\documentclass[10pt,conference]{IEEEtran}
\IEEEoverridecommandlockouts
\usepackage{cite}
\usepackage{amsmath,amssymb,amsfonts}
\usepackage{algorithmic}
\usepackage{graphicx}
\usepackage{textcomp}
\usepackage{xcolor}

\usepackage{booktabs} 

\usepackage{caption,subcaption}
\usepackage{url}
\usepackage[misc,geometry]{ifsym}
\usepackage{tikz}
\usepackage[figure, linesnumbered,vlined]{algorithm2e}
\usepackage{multirow}
\usepackage{colortbl}
\definecolor{mygray}{gray}{.9}
\usepackage{color}

\def\BibTeX{{\rm B\kern-.05em{\sc i\kern-.025em b}\kern-.08em
    T\kern-.1667em\lower.7ex\hbox{E}\kern-.125emX}}
\begin{document}

\title{Optimizing seed inputs in fuzzing with machine learning~\thanks{This research was supported in part by National Natural Science Foundations of China (Grant No. 61471344, 61772506) and National Key R\&D Program of China (Grant No. 2017YFB0802902).} }

\author{
    \IEEEauthorblockN{Liang Cheng\IEEEauthorrefmark{1}, 
    Yang Zhang\IEEEauthorrefmark{1}, 
    Yi Zhang\IEEEauthorrefmark{2}, 
    Chen Wu\IEEEauthorrefmark{3}, 
    Zhangtan Li\IEEEauthorrefmark{1}, 
    Yu Fu\IEEEauthorrefmark{1} and Haisheng Li\IEEEauthorrefmark{4}} 
    
    \IEEEauthorblockA{\IEEEauthorrefmark{1}TCA Lab\\
    Institute of Software, Chinese Academy of Sciences, Beijing, China\\ Email: chengliang@iscas.ac.cn} 
    
    \IEEEauthorblockA{\IEEEauthorrefmark{2} Email: yzhang7874@gmail.com}
    
    \IEEEauthorblockA{\IEEEauthorrefmark{3}diDST NLP, Alibaba, Hangzhou, China\\ Email: wuchen.wc@alibaba-inc.com} 
    
    \IEEEauthorblockA{\IEEEauthorrefmark{4}Beijing Technology and Business University, Beijing, China\\ Email: lihsh@th.btbu.edu.cn}
}

\maketitle

\begin{abstract}

The success of a fuzzing campaign is heavily depending on the quality of seed inputs used for test generation.  It is however challenging to compose a corpus of seed inputs that enable high code and behavior coverage of the target program, especially when the target program requires complex input formats such as PDF files. We present a machine learning based framework to improve the quality of seed inputs for fuzzing programs that take PDF files as input. Given an initial set of seed PDF files, our framework utilizes a set of neural networks to 1) discover the correlation between these PDF files and the execution in the target program, and 2) leverage such correlation to generate new seed files that more likely explore \emph{new} paths in the target program. Our experiments on a set of widely used PDF viewers demonstrate that the improved seed inputs produced by our framework could significantly increase the code coverage of the target program and the likelihood of detecting program crashes.    

\end{abstract}


\section{Introduction}
Fuzzing has been widely used to detect security vulnerabilities and bugs in IT systems because of its high efficiency. Most existing fuzzing tools, or \emph{fuzzers}, generate excessive test inputs by mutating a pre-selected corpus of seed  inputs with the hope to reveal potential bugs in the target program.  Therefore, extensive research effort has been dedicated to improving the quality of seed corpora~\cite{godefroid_learn&fuzz:_2017}.  Existing approaches in this direction, however, share a common limitation that they focus on discovering syntactic or semantic constraints posed by the target program for inputs in order to generate valid seed inputs.  As a result, seed corpora generated by these approaches often include too many redundant seed inputs that waste fuzzing effort by triggering the same execution paths in the target program. 

\begin{figure}\center
    \includegraphics[width=0.45\textwidth]{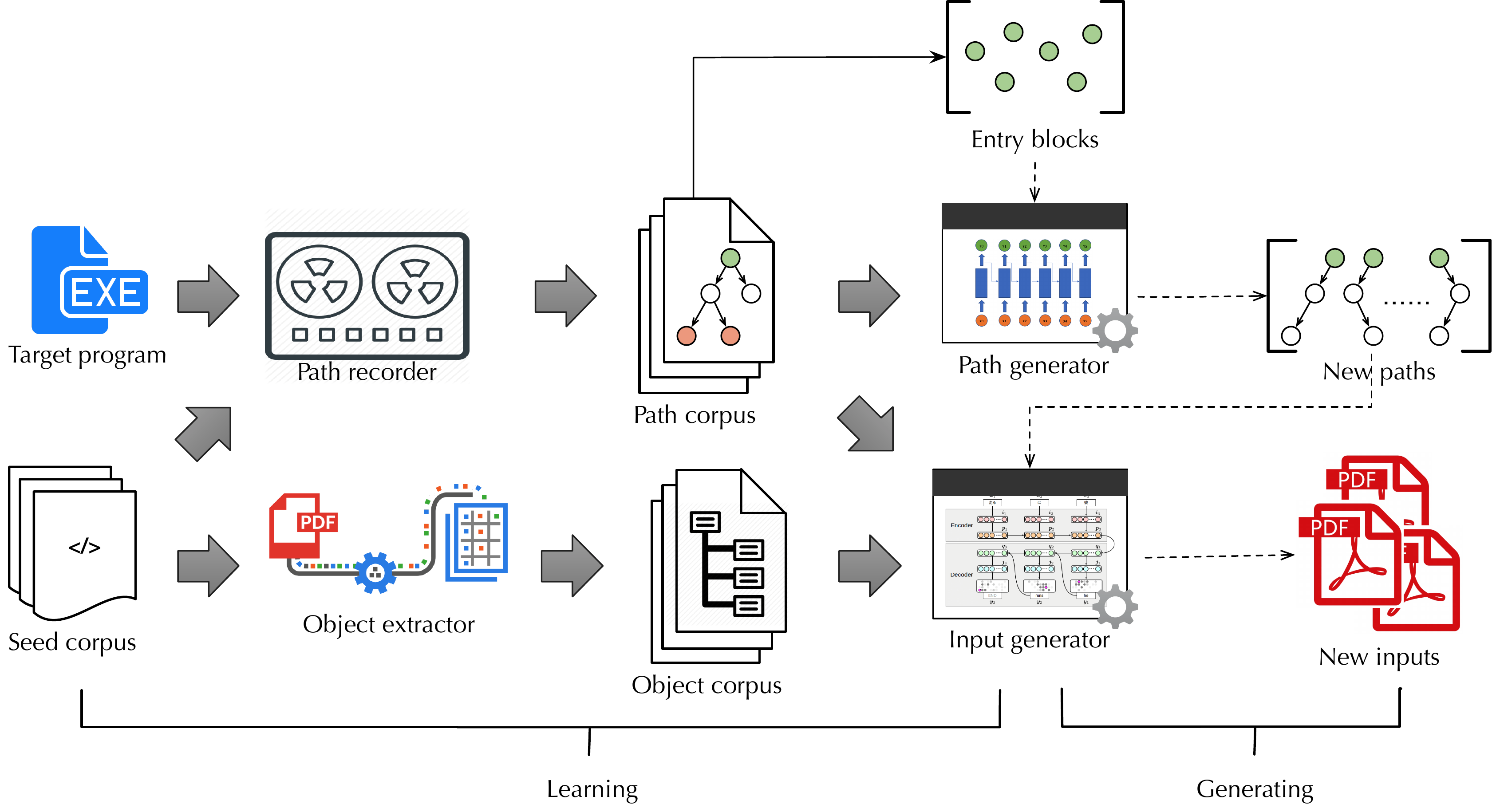}
    \caption{A framework for improving seed inputs in fuzzing.}
    \label{fig:archi}
\end{figure}

To address this limitation, we present a machine learning based framework that discovers and leverages the correlation between seed inputs and the execution of the target program to generate new seed inputs that trigger higher code coverage of the target program (and hence increase the chance of bug/crash detection) than the original seed inputs. Notably, our framework can work in combination with techniques that optimize the test mutation strategies in modern fuzzers (e.g., ~\cite{she_neuzz:_2018}) to further improve the effectiveness and efficiency of fuzzing.

Our framework first utilizes a generative model that bases on recurrent neural networks (RNNs) to generate new execution paths of the target program not covered by the original seed corpus. The new execution paths are then forwarded to  a \emph{sequence-to-sequence}(Seq2seq)-based transition model to translate into  valid PDF files (i.e., new seed inputs) triggering them.  In these tasks, both models are trained with the original seed inputs and corresponding execution of the target program. 

We have conducted a set of experiments on widely used PDF viewers, which demonstrates that new seed inputs produced by our framework significantly increased the code coverage of the target program and the likelihood of detecting program crashes.  Additional experiments also confirmed that our framework is applicable to other input formats such as PNG and TTF files with minimal customization.

\section{A Seed Input Generation Framework}\label{sec:overview}

The presented framework, as illustrated in Figure~\ref{fig:archi}, generates new seed inputs in three steps:

\textbf{Step 1: Data Preparation.} The \emph{Path Recorder} in Figure~\ref{fig:archi}, built upon Intel's instrumentation tool Pin, first feeds the original seed corpus to the target program and records the resulting execution sequences.  These execution paths are encoded as the starting addresses of the basic blocks along the paths and stored in the \emph{path corpus}. Given that execution paths are often too lengthy to be handled by the RNN models, a path compression algorithm is introduced to compress long paths down to a length less than 300 by replacing short sequences of basic blocks shared by multiple execution paths with super-blocks. 

\textbf{Step 2: Path Generation.} Execution paths in the \emph{path corpus} are used to train the \emph{Path Generator}, an RNN-based language model built on top of Andrej Karpathy's Char-RNN implementation\footnote{https://github.com/karpathy/char-rnn}, in order to learn the conditional distribution of basic blocks on these paths. This language model inherits the two-layer structure of standard char-RNNs, one for learning how basic blocks form functions and the other for learning how functions form complete execution paths, where the number of hidden states in each layer is set to 256. 

When queried with an initial basic block, the fully trained Path Generator is able to generate the rest of an execution path that has not been covered by previous execution paths (including those in the path corpus).  Two sampling strategies, \emph{Sample} and \emph{SampleFunction}, are introduced to the Path Generator to ensure the diversity of the generated execution paths. Under these strategies, the Path Generator samples the learned distribution either when the next basic block is predicted or when the current basic block is at the end of a function, respectively. 

\textbf{Step 3: Seed Generation.} Execution paths produced by the Path Generator are 'translated' by the \emph{Input Generator} into PDF files (i.e., new seed inputs) that trigger these execution paths. The Input Generator includes: 1) an \emph{Object Extractor} that retrieves all object sequences from PDF files in the original seed corpus; and 2) a Seq2Seq model that, after being trained with the path corpus and the corresponding object sequences retrieved by the Object Extractor,  learns and leverages the correlation between the original seed corpus and the path corpus to achieve an accurate translation from new execution paths to new seed inputs. 

The Seq2Seq model, implemented on top of the general-purpose Se2Seq framework\footnote{https://github.com/google/seq2seq}, includes an encoder RNN and a decoder RNN, where the size of both RNNs is set to 256, and the dropout rate is set to 0.5 for the former and 1 for the latter (to avoid potential overfitting issues).

\section{Evaluations}~\label{sec:eval}


\begin{table}  
\caption{Comparison of code coverage triggered by different seed corpora, where $C_{s}$ and $C_{sf}$ denote improved seed corpora generated using the $Sample$ and $SampleFunction$ strategies, respectively.}
\label{tab:corpus_cov}
\begin{tabular}{lccccc}
\toprule
    & \multirow{2}{*}{Original} & \multicolumn{2}{c}{$C_{s}$} & \multicolumn{2}{c}{$C_{sf}$} \\
    \cmidrule{3-4}\cmidrule{5-6}
                    & &  \#  & \%      &  \#  &  \%  \\
    \midrule
    basic blocks    & 4,548  & +113  & +2.48\% & +109  & +2.40\%  \\
    execution paths & 14,522 & +1008 & +6.94\% & +3528 & +24.30\% \\
    \bottomrule
  \end{tabular}
  \vspace{-6pt}
\end{table}

We evaluated our framework against the widely-used PDF viewer MuPDF:  a total of 43,684 PDF files (5.2 GB in size) were first downloaded from the Internet and fed to  MuPDF (using AFL - one of the most used greybox Fuzzers\footnote{http://lcamtuf.coredump.cx/afl/}).  Execution paths thus acquired and the downloaded PDF files were used to train our framework for 24 hours (on a computer with 4-core Intel i7-7700 CPU, 16G RAM and a NVidia GTX 1080 Ti GPU).  Table~\ref{tab:corpus_cov} shows that the new seed corpora generated by our framework caused up to 2.48\% more basic blocks and 24.30\% more execution paths being covered than the original seed corpus. Our results significantly surpassed similar works such as \cite{godefroid_learn&fuzz:_2017}, which generated seed corpora by learning the grammar of the PDF files and the new corpora covered 0.11\% more instructions.   

We next evaluated our framework by fuzzing MuPDF and three other PDF viewers (pdfium, podofo, and poppler) with the original and generated corpus for 24 hours. This produced similar results: the improved seed inputs generated by our framework explored on average 23.21\% more basic blocks and 31.69\% more execution paths. In addition, the improved seed inputs triggered 67 crashes in the PDF viewers under fuzzing including 2 CVE vulnerabilities, as compared to only 32 crashes (with none CVE vulnerability) triggered by the original seed corpus.

We also applied our framework to \emph{libpng} (a PNG reference library) and \emph{freetype} (an open source TTF library). Trained with PNG and TTF files extracted from the downloaded PDF files, our framework generated new seed inputs that led to significant code coverage increase in both target programs (i.e., 53.90\% more paths and 22.38\% more edges covered in freetype after 24-hour fuzzing). This might suggest that our framework is applicable to other complex input formats.

\bibliographystyle{IEEEtran}
\bibliography{./fuzzing-new}

\end{document}